\documentclass[a4paper,twocolumn,DIV=16,9pt]{scrartcl}

\usepackage[normalem]{ulem}

\usepackage[utf8]{inputenc}
\usepackage{graphicx}
\usepackage{amsmath}
\usepackage{url}
\usepackage{hyperref}
\usepackage{cleveref}
\usepackage{microtype}
\usepackage{abstract}
\usepackage{float}
\usepackage{titling}
\usepackage{enumitem}
\usepackage[firstinits=true,backend=bibtex]{biblatex}
\usepackage{bold-extra}
\usepackage{xspace}
\usepackage{xcolor}
\usepackage[textsize=footnotesize,obeyDraft,color=yellow]{todonotes}
\usepackage[all]{nowidow}
\usepackage[binary-units]{siunitx}
\usepackage{subfigure}
\newfloat{acks}{b}{lop}
\setlist[itemize]{itemsep=0.0em, topsep=0.5em, partopsep=0pt, leftmargin=1em}

\newcommand{\slicekit}{\mbox{SLICEKIT-A16}\xspace}
\newcommand{\xsonel}{\mbox{XS1-L}\xspace}

\SetLabelAlign{parright}{\parbox[t]{\labelwidth}{\raggedleft#1}}

\graphicspath{{./}}

\bibliography{bib}

\begin{document}

\setlength{\droptitle}{-1em}
\pretitle{\begin{center}\sffamily\bfseries\huge}
\title{Modeling and visualizing networked multi-core\\ embedded software energy
consumption}
\posttitle{\end{center}}

\date{Technical Report, August 2015}
\author{
    Steve Kerrison and Kerstin Eder\\
    University of Bristol, United Kingdom\\
    \texttt{firstname.lastname@bristol.ac.uk}
}

\makeatletter
\twocolumn[
  \begin{@twocolumnfalse}
    \maketitle
    
    \begin{abstract}

      In this report we present a network-level multi-core energy model and a
      software development process workflow that allows software developers to
      estimate the energy consumption of multi-core embedded programs.
      This work focuses on a high performance, cache-less and timing
      predictable embedded processor architecture, XS1. Prior modelling work is
      improved to increase accuracy, then extended to be parametric with
      respect to voltage and frequency scaling (VFS) and then integrated into a
      larger scale model of a network of interconnected cores. The modelling is
      supported by enhancements to an open source instruction set simulator to
      provide the first network timing aware simulations of the target
      architecture. Simulation based modelling techniques are combined with
      methods of results presentation to demonstrate how such work can be
      integrated into a software developer's workflow, enabling the developer
      to make informed, energy aware coding decisions. A set of single-,
      multi-threaded and multi-core benchmarks are used to exercise and
      evaluate the models and provide use case examples for how results can be
      presented and interpreted. The models all yield accuracy within an
      average $\pm\SI{5}{\percent}$ error margin.

    \end{abstract}

    \vspace{2.0\baselineskip}
  \end{@twocolumnfalse}
]
\makeatother

\begin{acks}
  \noindent\rule{\columnwidth}{0.4pt}
  \small
 
  The research leading to these results has received funding from the European
  Union Seventh Framework Programme (FP7/2007-2013) under grant agreement no
  318337, ENTRA - Whole-Systems Energy Transparency.

\end{acks}

\section{Introduction}
\label{sec:intro}

An increasing number of embedded systems now express their workloads through
concurrent software. The parallelism present in modern devices, in forms such
as multi-threading and multiple cores, allow this concurrency to be exploited.
This progression towards parallel systems has two main motivations. The first
is in response to hitting operating frequency limits, where more work must now
be done per clock in order to achieve performance gains in each new device
generation. The other uses parallelism to allow work to be completed on time at a lower operating frequency, which can yield significant energy reductions.

However, parallel systems and concurrent software introduce complexities over
traditional sequential variants that simply valued ``straight-line speed''. In
particular, synchronisation of and exchanging information between concurrent
components can negatively impact parallel performance if done inefficiently as
per the well known \emph{Amdahl's Law}. A good understanding of the software's
behaviour, coupled with appropriate underlying hardware can overcome this if
used correctly.

In embedded systems software, predictability is essential, both in terms of
execution time, where real-time deadlines must be met, and in terms of energy
consumption, where the supply of energy may be scarce. Time and energy are
related through power, and while significant effort is put into timing
predictable software, there remains both a lack of intuition and a lack of
tools to help software developers determine the energy consumption of their
modern embedded software components.

This report presents an energy model for a family of cache-less,
time-deterministic, hardware multi-threaded embedded processors, the XMOS
\xsonel series, which implements the XS1 architecture. These processors are
programmed in a C-like language with message passing present in both the
architecture and the programming model. The processors can be assembled into
networks of interconnected cores, where the communication paradigm then extends
across this network. The energy model must therefore be able to account for
software energy consumption within each core as well as the timing and power
effects of network traffic. To achieve this and also give developers better energy estimation tools, the following contributions are made:
  
\begin{itemize}

  \item A multi-threaded energy model for the
    \xsonel~\cite{Kerrison:2015:EMS:2764962.2700104} is extended to include
    more accurate instruction energy data, through greater instruction
    profiling and regression tree techniques.

  \item Support for Voltage and Frequency Scaling (VFS) is integrated into the
    model, the provide a richer environment for design space exploration by
    software developers.

  \item Several new features are added to \texttt{axe}, an Instruction Set
    Simulator (ISS) for the \xsonel, improving its core timing accuracy and
    introducing network timing behaviour, which has until now not been present
    in any simulators for these devices.

  \item The energy consumption of network communication is profiled, in order
    to extend the energy model to account for communication between
    multi-cores.

\end{itemize}

These contributions allow traces from the \texttt{axe} ISS to be analysed by
the modelling framework, producing both text reports and visualisation of
energy consumption across the network of processors in the system. The accuracy
of this work is established through a series of multi-threaded, multi-core
embedded software benchmarks. These are used to evaluate the effectiveness of
the modelling and detail how it can be used to aid a developer's design and
implementation decisions.

Results show that the core model average error is \SI{2.67}{\percent} with a
standard deviation of \SI{4.40}{\percent}, improving upon the prior work. The
network capable model demonstrate an average error of \SI{-4.92}{\percent},
with a standard deviation of \SI{3.92}{\percent}, supported by the VFS model
with a mean squared error of \SI{2.60}{\percent} and total error range of
\SI{15.72}{\percent}. The network model is shown to be suitable for determining
the best approach for implementing two concurrent signal processing tasks on a
target dual-core \xsonel platform.

\subsubsection*{Structure}

The rest of this report is structured as follows. Related work is presented in
\Cref{sec:related}, which looks at energy modelling of modern embedded
processors, multi-core communication techniques and parallelism in embedded
architectures, and summarises the particular implementations used in the \xsonel
processor. The core- and network-level energy models are explained in
\Cref{sec:model}, then the necessary ISS changes to support the model are
presented in \Cref{sec:axe}. Results from benchmarks exercising various parts
of the model and simulation framework are discussed in \Cref{sec:results} along
with an evaluation of their performance in terms of accuracy and usability.
Finally, \Cref{sec:conclusions} draws conclusions from this research and
proposes future work.

\section{Related work and background}
\label{sec:related}

Energy modelling of software is motivated by a need to reduce global ICT energy
consumption as well as to enable devices such as embedded systems to provide
more features and last longer on limited source of energy. Although hardware
actually consumes energy, it does so at the behest of software, which can be
inefficient if the software does not fit well to the target hardware, or does
not allow the hardware to exploit its own energy saving features~\cite{Roy1997}.

Multi-core systems have proliferated through ICT, from servers in datacenters
down to smart phones, and now even deeply embedded systems. Any endeavour to
provide software energy consumption metrics must therefore be multi-core aware.
In the rest of this section we discuss related work in three background areas.
First is multi-core processors in embedded systems, next is energy modelling of
processors, with a focus on software level energy consumption, and finally we
introduce the \xsonel processor, the particular micro-architecture used as a
case study for this research.

\subsection{Parallelism and multi-core embedded processors}

There are various ways of realising parallelism in processors. In embedded
systems, many methods have been used. VLIW (Very Long Instruction Word) has
been used for some time, particularly in DSPs (Digital Signal Processors),
where instruction packets enable software pipelining to be
parallelised.Multi-core is becoming more prevalent, where it is beneficial to
replicate a core several times and distribute work between cores. This has
become necessary to provide performance gains as frequency increases have
become harder to realise within practical power budgets~\cite{itrs2013}.

High performance embedded processors, such as those found in smart phones, can feature multiple cores with different micro-architectures. ARM's \emph{big.LITTLE} is the seminal example of this, where programs can be scheduled onto simpler cores when low-energy operation is necessary or appropriate. The \emph{little} cores are slower, but can be operated at a lower voltage and frequency point than their \emph{big} counterparts, consuming significantly less energy. In big.LITTLE, significant effort is put into cache coherency between the cores, and migrating tasks can require flushing and copying of core-local caches in order to keep consistent state.

Smaller processors, such as ARM's Cortex-M series, can also be used in
multi-core, but the implementation is defined by the manufacturer. ARM has made
recommendations on how to construct such devices, including cache and memory
arbitration mechanisms~\cite{M-series-multicore}. Older generation ARM9
processors have been assembled in their thousands in the \emph{SpiNNaker}
system~\cite{Furber2013}.

Rather than connecting processors via a cache hierarchy and memory bus, some
systems implement a network of cores. Devices such as the Adapteva
Epiphany~\cite{epiphanyIntro} and EZChip TILE~\cite{TileGx} processors feature
many cores in a Network-on-Chip, with a grid topology of interconnects between
them. In both of these processors there are multiple networks, each serving a
unique purpose, such as I/O, cache coherency and direct inter-tile
communication. The Intel Xeon Phi~\cite{XeonPhi2013} uses a ring network and a
hierarchy of processors, caches, tag directories and memory controllers to
create a NoC that can also be viewed as a traditional memory hierarchy. Its use
is not in embedded systems, but rather as a high performance computing
accelerator.

The \xsonel processor features no cache hierarchy and can be assembled into a network of cores where channel style communication is possible both on- or off-chip. This is discussed in more detail in \Cref{sec:xs1-l}.

\subsection{Energy modelling of processors}

A program's energy consumption is the integral of a device's power dissipation
during the course of execution:
\begin{equation}
  E = \int_{t=0}^{T} P(t)\ dt \text{,}
  \label{eq:energy_integral}
\end{equation}
although this is frequently represented using an average power, giving $E = P
\times T$. To energy model a processor, $P$ must be estimated over the course
of $T$ with sufficient granularity and precision to provide a desired accuracy.
At the hardware level, detailed transistor or CMOS device models can be used,
and every change in circuit state simulated to determine a fine-grained power
estimation. This is time consuming and requires access to the RTL description
of a processor, making this form of analysis infeasible for software
developers.

Higher level models can be used instead, such as those modelling the processor as functional blocks. Instructions issued by the processor trigger activity in the functional blocks, and a cost is associated with that, which can be used to estimate the energy consumption of a sequence of instructions. At this level, the instructions are an essential part, as these drive the modelling, but also form a connection to the software --- the instruction sequences for a given architecture are related to the software developer's program via transformation by a compiler. The ISA therefore provides a good level at which to perform analysis of hardware energy consumption at the behest of software.

Seminal work in ISA level energy modelling includes that of
Tiwari et al.~\cite{Tiwari1994a}, where sequences of instructions are assigned
costs, as well as the transitions between instructions, which causes circuit
switching as new control paths are enabled. This work has been drawn up upon to
enable energy consumption simulation frameworks such as
Wattch~\cite{Brooks2000} and SimPanalyzer~\cite{Simpanalyzer}. This style of
ISA level modelling has also been refined to include finer grained detail on
the activity along the processor data path, where data value changes also
influence energy consumption~\cite{Steinke2001}.

Energy modelling has been performed for a wide variety of processors with
various micro-architectural characteristics, for example VLIW DSP
devices~\cite{Ibrahim2008a}, both simple and high performance ARM variants, as
well as very large processors such modern server grade x86
devices~\cite{sniperMcPAT2012} and the 61 core Xeon Phi~\cite{phimodel}. These
all draw from similar background, but account for different processor features,
and obtain their model data from different sources. For example, high
performance ARM and x86 models can use hardware performance counters to model
activities such as cache misses, which have a significant impact on energy
consumption. Simpler devices may not be so affected, and thus direct
instruction level costs can be attributed. Parametrised energy models that
consider properties such as operating frequency and voltage have been created
for other processors, such as the Intel Xeon in~\cite{Bartolini2011}, to inform
a model predictive controller in order to smooth thermal hotspots in such dense
multi core devices.

A single core model of the XMOS \xsonel architecture is presented in~\cite{Kerrison:2015:EMS:2764962.2700104}, which uses data from a series of instruction energy profiling tests in order to build the model. The architecture's hardware multi-threading is accounted for, with the level of parallelism (active threads) contributing to energy consumption during the course of the analysed program. This model has been applied using instruction set simulation, and also via static analysis at the ISA level~\cite{grech15,isa-energy-lopstr13} as well as the LLVM IR level~\cite{kyriakosWCEC}.

\subsection{The \xsonel processor and network}
\label{sec:xs1-l}

The \xsonel family is a group of processors implementing the XMOS XS1 ISA in a
\SI{65}{\nano\meter} process technology, featuring a configurable network upon
which arbitrary topologies of interconnected processors can be built. Each core
has a four stage pipeline and support for up to eight hardware scheduled
threads. A thread can have no more than one instruction in the pipeline at any
given time, therefore the \xsonel parallelism is only fully utilised if four or
more threads are active.

These processors include \SI{64}{\kibi\byte} of single cycle SRAM and have no
cache, therefore the memory subsystem is flat and requires no special
considerations with respect to timing. The majority of instructions complete in
four clock cycles, with the exception of the divide and remainder instructions
and any instructions that block on some form of I/O. If more than four threads
are active, then the instruction issue rate per thread will reduce
proportionally, but the instruction throughput of the processor remains the
same. This makes timing analysis of the processor very predictable, allowing
tight bounds or even exact values to be produced.

\begin{figure}
  \centering
  \includegraphics[width=1.0\columnwidth, clip, trim=0cm 10.5cm 9cm 0cm]{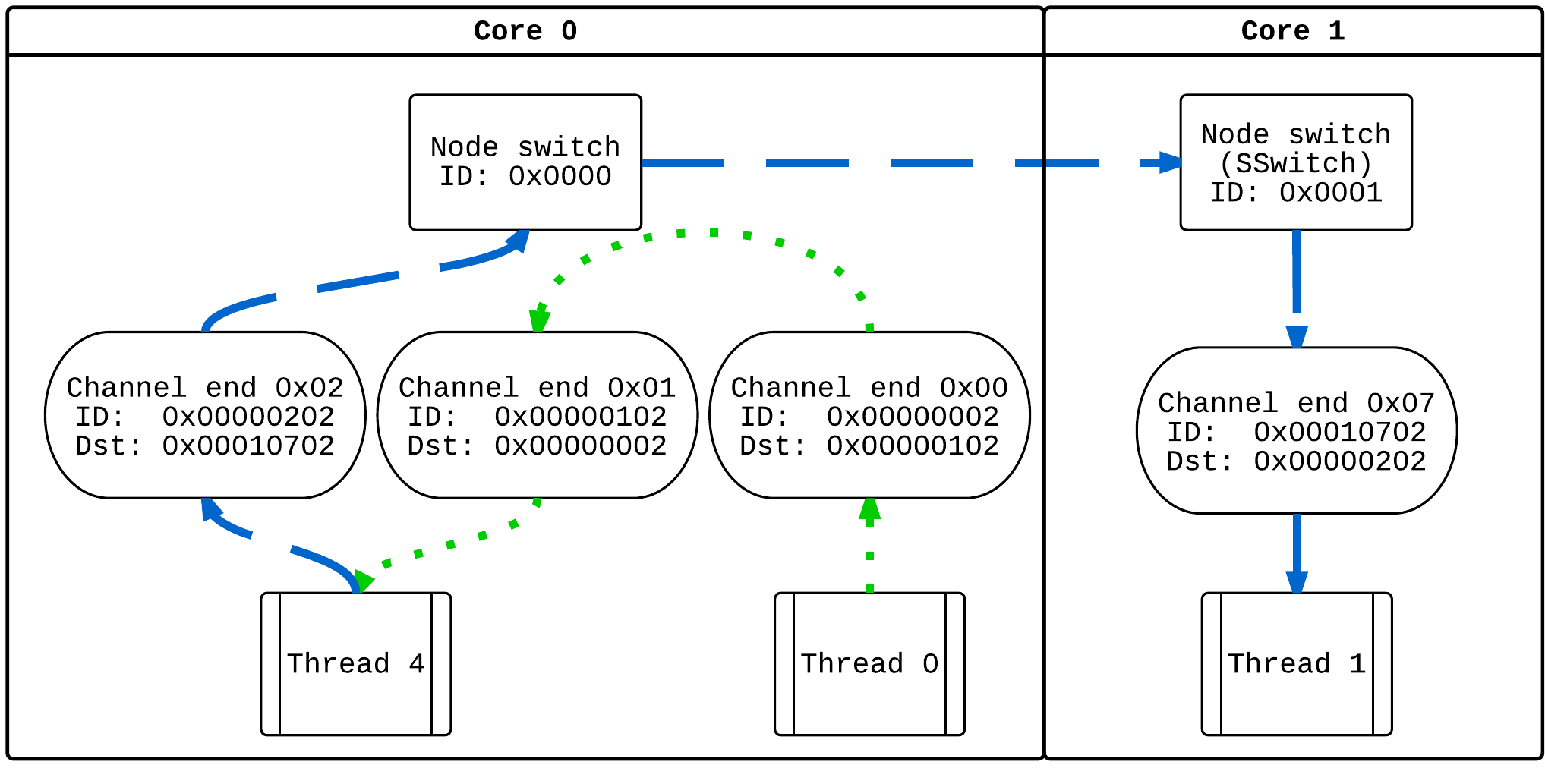}
  \caption{Visualisation of channel based communication between threads both locally and between cores.}
  \label{fig:chancomm}
\end{figure}

The XS1 instruction set includes provisions for \emph{resource operations}.
These are interactions with peripheral devices, such as I/O ports,
synchronisers and communication channel endpoints (chanends). As such,
activities such as I/O are a first class member of the instruction set. Other
instruction sets, such as x86, have similar
provisions~\cite[pp.115,176]{x86manual}. However, the XS1 architecture takes
this further, and places these peripherals outside of the memory space, such
that I/O and other resource operations are not translated into memory mapped
reads and writes, but are instead a completely separate data path. This
separation of memory and resources aids in the modelling processor,
particularly when communication between threads and cores is considered.

On a single core, it is possible for threads to communicate or access common
data using shared memory paradigms. This can be expressed in software through
appropriate use of regular pointers in C, or through specially attributed
pointers in version 2 of the \emph{XC} language that was developed to
complement XS1.  However, CSP style channel communication is more prevalent in
XC. Channels in XC translate into channel endpoints in the XS1 architecture,
where two chanends are logically connected together. Communication then takes
the form of \texttt{in} and \texttt{out} instructions. Control tokens can be
used to provide synchronisation, and instructions will block if buffers are
full or no data is available to read. This paradigm extends beyond core-local
communication and out onto a network of cores. Therefore, concurrent programs
can grow to use multiple processors with relative ease.

A network of \xsonel processors consists of multiple cores each connected to
their own integrated switch. This switch provides a number of links, which can
be connected to other switches, either on- or off-chip. These links can operate
in either five- or two-wire mode in each direction, where the former can carry
two bits per symbol and the latter one bit per symbol in an 8b/10b encoding.
The five wire mode is therefore faster at the same frequency, but requires ten
wires total per link. Each link possesses a receive buffer and credit based
flow control is used to prevent overrun. When a link is first enabled, the
sending switch must solicit credit from the switch at the other end of the link
with a \emph{hello token}. During normal operation, \emph{credit} tokens are
sent from the receiver to the sender as buffer space becomes available.

Routing between switches is configurable based on IDs assigned to each node,
where a node is a switch and its associated core. When a message is first sent
from the chanend of a core, the ID of the destination node is prepended to the
message. Receiving switches then compare this ID to their own. If they are
different, the first bit that is different is used to determine the
\emph{direction} along which the message will be routed. A direction can be
assigned one or more links, and the next available link in that direction will
then be used for forwarding. Typically, dimension-order routing is used to
create a deadlock avoiding network, but this is dependent upon the topology
network that is physically assembled. Links are held exclusively by the source
chanend either indefinitely, or until a closing control token is transmitted
from the source. Through this approach, both wormhole routed packets and
permanently reserved streaming routes can be created.

A high level view of threads communicating through chanends and switches, both
locally and between cores is shown in \Cref{fig:chancomm}.  The precise
implementation details and configuration parameters are detailed
in~\cite{xs1architecture,XS1Lsys2008}.  Examples of multi-core XS1
implementations include the XMP-64, which features 64 cores, using the older
XS1-G family, and the Swallow project, which assembles multiple dual-core
\xsonel family processors into a system of hundreds of
cores~\cite{swallowOverview}.  These use hypercube and lattice network
topologies, respectively.

\section{\xsonel core and network energy model}
\label{sec:model}

In this report, the modelling effort of~\cite{Kerrison:2015:EMS:2764962.2700104}
is extended in several ways.  Firstly, more instructions are directly energy
profiled, and for those that cannot, a regression tree approach is implemented
to estimate their energy cost. Secondly, additional voltage and frequency
profiling is performed, using a suitable variant of the \xsonel, to produce a
VFS aware model version, retaining good error bounds. Finally, network
communication costs are considered, through further profiling, and a network
level, communication ware model produced, integrating core, switch and
interconnect components within the model.

\subsection{Regression tree}

\begin{figure}
  \includegraphics[width=1.0\columnwidth]{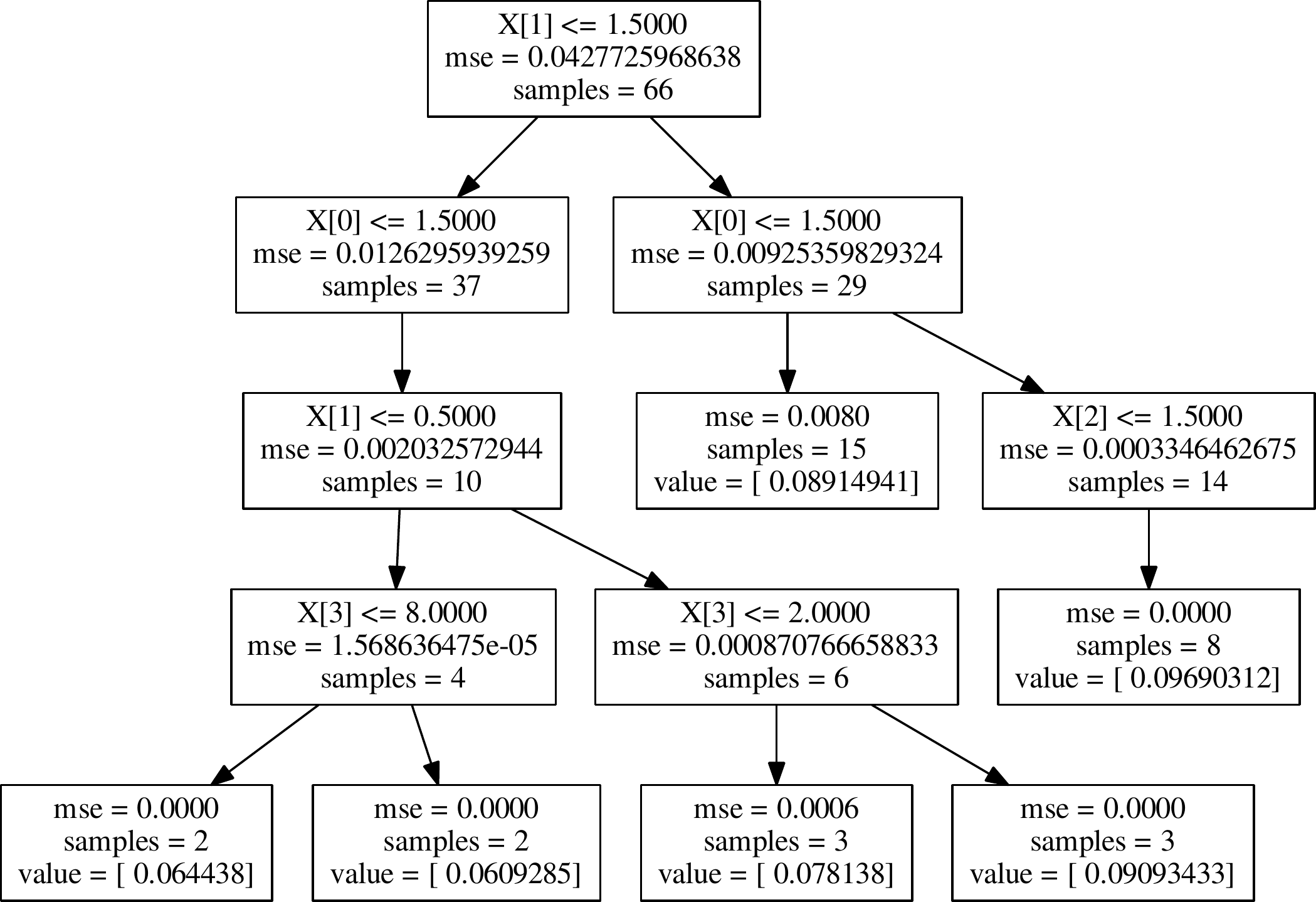}
  \caption{Visualisation of part of the model regression tree. Leaves provide energy estimations, all other nodes are decisions based on a particular instruction feature $X[f]$. Not all branches are shown; the full tree is 29 nodes.}
  \label{fig:rtree}
\end{figure}

The prior work of~\cite{Kerrison:2015:EMS:2764962.2700104} investigated
grouping instructions by operand count in order to provide an energy estimate
for un-profiled instructions, as well as to reduce model complexity. However,
the evaluation showed that this was not suitably fined grained or sufficiently
accurate. Instead, each profiled instruction is accounted for individually, and
un-profiled instructions are assigned a \emph{default} value, based on the
observed average of all profiled instructions.

Here, a different approach is used, where a set of instruction features are
used to classify each instruction. This is combined with the direct instruction
profiling data into a regression tree, allowing un-profiled instructions to
receive an energy estimate based on profiled instructions with similar
features.

\subsubsection{Tree construction}

First, a set of features are identified, which from empirical data, demonstrate a correlation with energy consumption. These are specific to the \xsonel processor, although can be re-defined for other processors in order to re-use the technique. In the case of the \xsonel, the features are:

\begin{description}[labelwidth=1em, align=parright]

  \item[L:] Instruction length (short or long: 1 or 2).

  \item[S:] Number of source registers (count: 0--4).

  \item[D:] Number of destination registers (count: 0--2).

  \item[I:] Length of immediate operand (num. bits: 4--16).

  \item[M:] A memory operation is performed (Boolean).

  \item[R:] A resource operation is performed (Boolean).

\end{description}

The \emph{Scikit-Learn} \texttt{DecisionTreeRegressor}~\cite{skikit-decision}
is used to build the regression tree. The data is presented as an matrix of
instruction features and a vector of measured energy for each profiled
instruction. From this, a regression based decision tree is constructed. A
sample of the input data is provided in \Cref{tab:regress-data}.

The regression tree construction library uses floating point feature
parameters. For the given feature set, both the integer and Boolean features
can be converted to their nearest floating point equivalent without
consequence.

\begin{table}
  \centering
  \begin{tabular}{|*{8}{c|}}
    \hline
    & \multicolumn{6}{c|}{\textbf{Features}} & 
    \\ \cline{2-7}
    \textbf{Instr.} & \textbf{L} & \textbf{S} & \textbf{D} & \textbf{I} & \textbf{M} & \textbf{R} & \textbf{Energy}
    \\ \hline
    \texttt{add\_3r} & 1 & 2 & 1 & 0 & 0 & 0 & \SI{185}{\milli\watt}
    \\ \hline
    \texttt{ldc\_lru6} & 2 & 0 & 1 & 10 & 0 & 0 & \SI{160}{\milli\watt}
    \\ \hline
    \texttt{outct\_rus} & 1 & 1 & 0 & 4 & 0 & 1 & \SI{134}{\milli\watt}
    \\ \hline
  \end{tabular}
  \caption{Input data for regression tree constructor}
  \label{tab:regress-data}
\end{table}

\subsubsection{Tree traversal}

A cutting of the regression tree for our energy model is depicted in
\Cref{fig:rtree}. When the energy cost of an instruction must be determined, a
check first determines if a direct energy measurement exists. If so, it can be
used within the model equation. If not, then the instruction's features are
used to traverse the decision tree. Each feature is indexed numerically by the
\texttt{DecisionTreeRegressor}, and map to the features in the order we have
declared them.

For example, the first decision, at the root of tree, is dependent upon the
number of source operands. Those with fewer than two (or $\le 1.5$) follow the
left branch, whilst those with two or more follow the right branch. The
instruction length is considered next. However, descending into the tree
further, the feature selection that minimises the mean squared error (mse),
will differ depending on the instruction and the collected energy data. This
makes the decision tree more versatile than a flat ordinary least squares
regression. For example, there are no instructions with four source operands
that use memory, therefore $X[4]$ or feature \textbf{M} has no influence upon
such instructions. The tree is also unbalanced; some branches reach leaves in
fewer levels, due to no variation in features beyond a certain decision point.

The accuracy of this approach is tested and evaluated in \Cref{sec:results},
where a reduction in both error and variance is shown when compared to the
previous model.

\subsection{VFS modelling}
\label{sec:vfsmodelling}

The \xsonel series of processors can dynamically adjust their core clock
frequency when idle, and some devices support variable voltage. The low speed
of voltage adjustment makes dynamic voltage and frequency scaling (DVFS)
impractical for most of the real-time embedded tasks targeted to the \xsonel.
However, it is still possible to statically select a best voltage and frequency
for a given set of tasks, and so there is motivation to provide an energy model
can support this exploration in order to determine what savings can be made.

An XMOS \emph{\slicekit} is used for VFS profiling, a board containing an
XS1-A16 processor. The A16 contains two \xsonel family processor cores, as well
as an analogue component block containing components such as ADCs and most
importantly configurable DC-DC power supplies, one of which services the cores.
The \slicekit board provides two shunt resistors for power sensing, one for the
\SI{3.3}{\volt} I/O used by the chip and one for the \SI{3.3}{\volt} supply fed
to the on-chip voltage regulators. Measurements are performed with a MAGEEC
power measurement board~\cite{mageecwand}, which provides 2\,MSPS at 12-bit
resolution with a noise floor at approximately \SI{0.1}{\percent} of measured power in our test setup, depending on the current
supplied. The measurement setup and power supply structure is different to that
used in~\cite{Kerrison:2015:EMS:2764962.2700104}, so the power supplies must be
considered in the new model.

\subsubsection{Profiling method}
\label{sec:vfsparams}

VFS profiling is performed by a series of tests both at idle and high power, where each test is performed at different voltage and frequency operating points. Three configurable parameters are exercised:
\begin{description}
  \item[System frequency] The frequency produced by the PLL, which is integrated into each processor's switch. This sets the frequency of the switch and core, each of which can then be divided further.
  \item[Core divider] The divider applied to the system frequency to produce the core frequency. Typically this is zero. The specified divider can be applied dynamically when there are no active threads, or permanently.
  \item[Core voltage] The power supply to the device's cores is configurable in \SI{10}{\milli\volt} steps.
\end{description}

Other parameters, such as the reference clock, can be also be changed. However,
the reference clock is used for timing ports and other synchronisation
activities, thus we keep it at its default of \SI{100}{\mega\hertz} to prevent
unexpected timing changes in programs. As a result of this, profiling is
limited to system frequencies of 500, 400 and \SI{300}{\mega\hertz} and a core
divider in the range 0--3. Error free operation was achieved with a core voltage range of 0.85--\SI{1.15}{\volt}. However, the vendor only certifies devices for operation at \SI{1.0}{\volt}, and extensive testing at each voltage point was not performed; they were used purely for VFS characterisation.

\subsubsection{VFS profiling data}

\begin{figure}
  \centering
  \includegraphics[width=1.0\columnwidth]{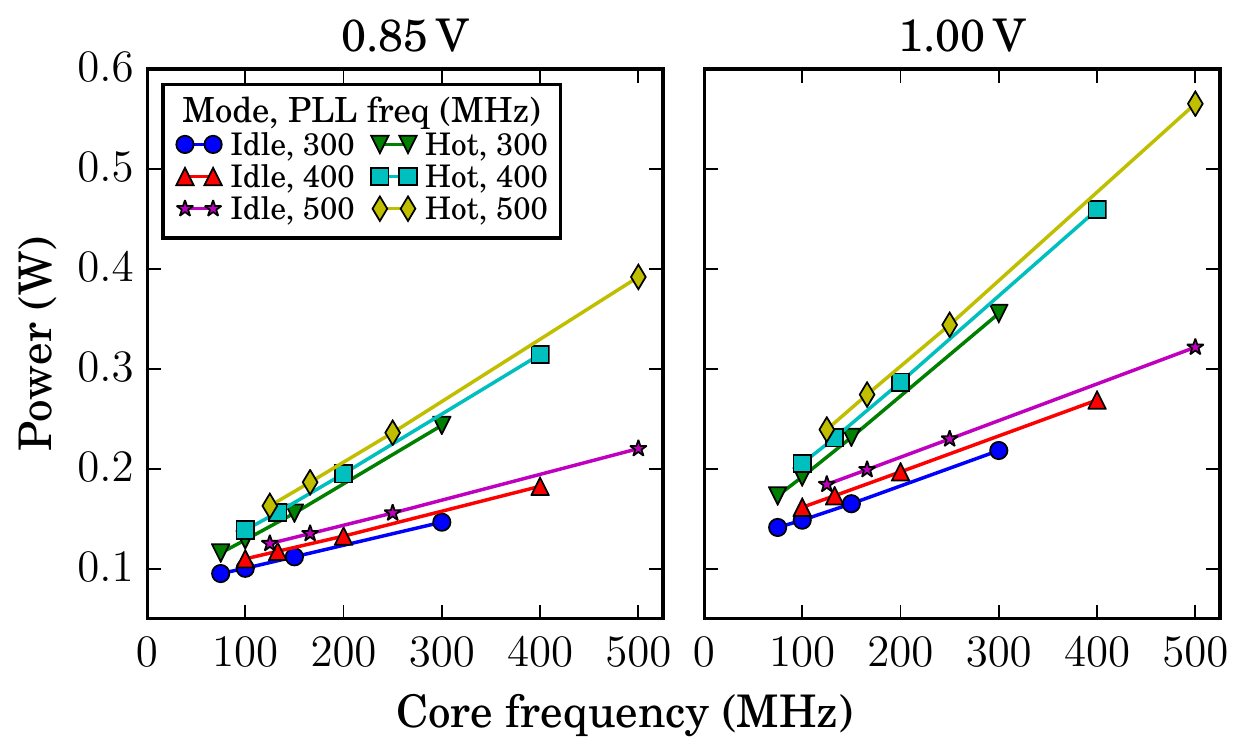}
  \caption{Power measurements at two voltage points for idle and high power
  tests over a range of system frequencies and dividers.}
  \label{fig:vfsdata}
\end{figure}

\Cref{fig:vfsdata} shows the profiling data for two of the voltage points that
were tested. Each plot shows six series; three for idle tests and three for
high power tests, each at one of three system frequencies. Points along the x
axis determine the core frequency after the divider is applied. 

The \SI{100}{\mega\hertz} operating point is achieved twice during tests, with
$F = \frac{400}{4}$ and $F = \frac{300}{3}$. From this we see that there is an
overhead in having a higher system clock, regardless of the resultant core
clock. This is intuitive, as there is still some part of the system operating
at the higher frequency.

\subsubsection{Energy model}

To produce a VFS capable energy model, we incorporate the configurable parameters defined in \Cref{sec:vfsparams} into a suitably modified model equation. Curve fitting is used to determine the contribution that these parameters have to the equation.

\emph{SciPy}'s Nelder-Mead method~\cite{SPNM} is used to minimize the error of
the function \Cref{eq:vfsmodelfunc} against the idle test profile data
collected, for the following parameters.  $C_\text{pll}$ is the characteristic
capacitance present at the system frequency (or PLL frequency). $C_\text{idle}$
is the characteristic capacitance in the core at idle. $I_\text{leak}$ is the
static leakage current.  Finally, $I_\text{ext}$ captures other effects
parametric to the supply voltage and scaled by the power dissipated in the
device, approximating power supply efficiency.
\begin{equation}
  F = (V^2C_\text{pll}F_\text{pll} + V^2C_\text{idle}F_\text{core} +
  VI_\text{leak}) \times VI_\text{ext}
  \label{eq:vfsmodelfunc}
\end{equation}
The resultant parameters are:
{
  \sisetup{
    round-mode          = figures,
    round-precision     = 3,
    scientific-notation = engineering
  }
  \begin{align}
    \nonumber C_\text{pll} &= \SI{6.7498630e-10}{} 
    ,& \quad C_\text{idle} &= \SI{1.6757538e-09}{} \\
    I_\text{leak} &= \SI{0.33368428}{}
    ,& \quad I_\text{ext} &= \SI{0.1060801}{}.
    \label{eq:vfsmodelparams}
  \end{align}

  A parameter is also determined for the high-power tests, $C_\text{hot} =
  \SI{2.1484374e-09}{}$, although it is used only for validation, and
  not in the final energy model. Testing these parameters against the profiling
  data, a mean squared error of \SI{2.6}{\percent} is achieved. The minimum
  error is $-3.58$\,\% and the maximum $12.14$\,\%, giving a full error range
  of $15.72$\,\%.

}
These parameters are then used in a modified version of the instruction
level energy model from~\cite{Kerrison:2015:EMS:2764962.2700104}. This yields
\Cref{eq:vfsmodel} as the new model:
\begin{align}
  \nonumber E_\text{instr} = &\ (\ V^2C_\text{pll}F_\text{pll} \\
  \nonumber & + V^2F_\text{core}(C_\text{idle}+C_\text{instr})M_{N_\text{pipe}}O \\
            & + VI_\text{leak}\ ) \times VI_\text{ext} \times 4T_\text{clk}
  \label{eq:vfsmodel}
  \\
  \nonumber \text{where} \quad M_\text{pipe} = &\ \min(4, N_t).
\end{align}

This captures the previous components of the model, plus the frequency and voltage dependent parameters. $N_\text{pipe}$ is the number of threads present in the pipeline when the instruction's energy is measured, which is the minimum of 4 and the number of active threads. This is used select a scaling factor due to parallelism, $M$. An average inter-instruction overhead, $O$, is also included, as per the original model.

The remainder of this report focuses more on network aware energy modelling,
rather than VFS design space exploration. Future work could include exercising
the VFS aspect of the energy model more heavily, and so is included here for
the benefit of such endeavours.

\subsection{Network modelling}

A system level model of a network of \xsonel processors is comprised of multiple
core model instances, as well additional modelling components to capture
network switch and link activity. The core model requires either simulated
instruction sequences or appropriately parametrised static analysis. A system level model must support the interaction between multiple cores. The implementation details of this at a simulation level, are covered in \Cref{sec:axe}.

\subsubsection{Parameters}

Communication costs must be accounted for in three system components. Firstly, the core, where the \texttt{in} and \texttt{out} instructions are executed. These are already captured in the core energy model. Second is the switch, which consumes energy as it routes tokens through it. Finally, the interconnects over which tokens are transmitted must be considered.

Switch energy consumption data is acquired from profiling of the larger Swallow
\xsonel system~\cite[p. 124]{KerrisonThesis}. Link energy for the
\slicekit is
determined from direct profiling. These are shown in terms of Joules per token
in \Cref{eq:commsmodel}.

{
  \sisetup{
    round-mode          = figures,
    round-precision     = 3,
    scientific-notation = engineering
  }
  
  \begin{equation}
    E_\text{switch} = \SI{70.83839e-12}{\joule}, \quad
              E_\text{link} = \SI{2.21e-10}{\joule}
    \label{eq:commsmodel}
  \end{equation}
}

Currently, these are fixed values. However, it is possible to parametrise these by link length (where longer wiring has a higher capacitance), as well as by switch frequency and voltage. This may form future work, joining well with the proposed further work on VFS modelling.

\subsubsection{Construction}

The network level model is constructed using the \texttt{NetworkX} library for
Python, which allows networks with nodes and edges that have arbitrary
attributes. The XML file used by the developer or vendor to describe an XMOS
based system (the XN file), is read by the energy modelling framework and used
to construct a graph of the system's cores, switches and links.

When a simulation trace is analysed by the modelling tool, energy is
incremented in each graph node or edge as appropriate. Instructions increase
the core energy of the relevant core, whilst token traces increase the source
switch and traversed link energy. For this trace analysis to account for
network activity, the trace must include network activity that identifies
tokens traversing links. This change was made to \texttt{axe} as part of the
modifications described in \Cref{sec:axe}.

At the end of the modelling run, this data can be aggregated into a text
report, broken down by core, or as a visualisation. These will be shown in
\Cref{sec:results}.

\section{ISS network and timing implementation}
\label{sec:axe}

\xsonel energy modelling has been demonstrated using statistics from instruction
set simulation as well as various levels of static analysis. Using full
instruction set simulation traces provide more detail, at the cost of
simulation time. However, by improving analysis of traces to complete once a
\emph{function or section of interest} has completed, simulation time can be
kept low. The same triggering methods used by the hardware measurement, can
easily be used to define sections of interest by identifying the relevant I/O
resource instructions in a trace. This means single iterations of functions or
algorithms can be observed by modelling, where repeated iterations are required
for physical measurement. The slowdown of simulation is mitigated to some
degree by this.  This is mitigated further by the use of \texttt{axe}, an open
source XS1 simulator that is faster than its closed source \texttt{xsim}
counterpart, although it can be less accurate. A number of \texttt{axe}
enhancements are detailed in this section that improve its accuracy, whilst
preserving some of its performance advantage.

Enabling full trace simulation allows better debugging of the energy model, as
well as the opportunity to more closely scrutinise where energy is being
consumed. To that end, this work focuses on full traces. However, the models
underpinning this work can be adapted for use at other levels of abstraction,
as with previous model versions.

\subsection{Instruction scheduling}

The modified version of \texttt{axe} enforces strict instruction scheduling,
where each active thread may only issue one instruction before the next queued
thread is given an opportunity. The timestamps of subsequent instructions in a
thread are incremented by $\min(4, N_t)$, to reflect the four-stage, hazard
free pipeline.

This more closely follows the micro-architecture, whereas the original
\texttt{axe} implementation may issue multiple instructions from one active
thread even when another thread is also in an active state. This also ensures
that the timestamps in instruction traces are ordered, greatly simplifying the
process of determining pipeline occupation during energy modelling.

\subsection{\texttt{\bfseries FNOP} simulation}

In addition to instruction scheduling changes, occurrences of fetch no-ops
(\texttt{FNOP}s) are also recorded in the modified simulator. A simple model of
the processor's instruction buffers is used to determine when a thread must
stall in order to fetch the next instruction word. The conditions leading to an \texttt{FNOP} include:

\begin{itemize}
  
  \item Sequences of memory operations in a thread, preventing any instructions
    being fetched for that thread during the memory stage of the pipeline.
  
  \item Branching to an unaligned 4-byte instruction, where only the first half
    of the instruction is fetched during the memory stage of the branch
    instruction.

\end{itemize}

Many \texttt{FNOP}s can be avoided by re-scheduling long instructions and
memory instructions amongst short, non-memory instructions, as well as word
aligning entry points to loops where the first instruction is long. However,
the compiler does not currently do all of these automatically. The impact of
\texttt{FNOP}s can be significant if they are present in tight loops,
increasing execution time and therefore energy. Thus, is it important to
correctly simulate this behaviour for accurate energy modelling.

\subsection{Switch and link control flow}

Both the \texttt{axe} and \texttt{xsim} \xsonel instruction set simulators
support channel communication in multi-core programs. However, even with
accurate core-local instruction scheduling, neither include accurate simulation
of network behaviour. At a functional level, link utilisation and route
reservation are simulated, such that protocol violations can be detected and
appropriate exceptions raised, as well as some degree of performance limiting
due to route contention. However, multi-core message tokens are transmitted in
zero time. This can create significant timing error when simulating communicating multi-core programs.

To address this, we have added a model of the link control flow mechanisms from \xsonel into \texttt{axe}'s switch, link and channel end code. Control tokens such as \texttt{HELLO} and the initial \texttt{CREDIT} issue~\cite[pp. 10--13]{XS1Lsys2008} are now handled rather than ignored. Symbol and token delays on links, as specified in the XN platform description file at compile time, are also obeyed. This ensures that messages traverse the network in a realistic time-scale, and that as buffers fill, network throughput is throttled.

\begin{table}
  \centering
  \begin{tabular}{|*{6}{c|}}
    \hline
    & \multicolumn{3}{c|}{\textbf{Recorded time (\SI{}{\micro\second})}} & \multicolumn{2}{c|}{\textbf{Error (\%)}}
    \\ \cline{2-6}
    \textbf{Test} & \textbf{HW} & \texttt{\bfseries xsim} & \texttt{\bfseries axe}  & \texttt{\bfseries xsim} & \texttt{\bfseries axe}
    \\ \hline
    One core & 2.270 & 1.118 & 2.272 & -50.75 & 0.09 
    \\ \hline
    Two core & 8.300 & 2.150 & 8.287 & -74.10 & -0.16 
    \\ \hline
  \end{tabular}
  \caption{1024-word message timing, comparing dual-core hardware to \texttt{xsim} and modified \texttt{axe} simulators.}
  \label{tab:commstiming}
\end{table}

The current changes are not completely faithful to the hardware, but yield a
significant improvement on the previous simulation capabilities. This is
evident in \Cref{tab:commstiming}, which compares a 1024-word transfer between
two channel ends on a \slicekit in both core-local and dual-core variants,
with respect to the actual hardware timing, \texttt{xsim} simulation and the
modified \texttt{axe} simulation. The error in \texttt{xsim} exceeds \SI{50}{\percent} in
both cases, whereas \texttt{axe} achieves less than \SI{0.2}{\percent}. Over a broader
range of similar tests, with different message lengths and producer/consume
rates, \texttt{axe} is able to maintain an average error of \SI{0.80}{\percent} with a
standard deviation of \SI{1.26}{\percent}.

To aid modelling, switch and link activity are added to \texttt{axe} simulation
traces. These resemble the switch tracing present in the vendor's \texttt{xsim}
simulator with the \texttt{--tracing-switch} parameter set, but are in a JSON
format that is more readily consumable by the energy modelling framework.

\section{Benchmarking and evaluation}
\label{sec:results}

This evaluation considers enhancements to the core model as well as the multi-core communication model. However, VFS modelling, as discussed in \Cref{sec:vfsmodelling}, remains for future work, due to the current instruction set simulation framework not supporting configurable operating frequencies and clock dividers without significant further development. This does not exclude the VFS model from use in other forms of non-simulation based analysis, however.

\begin{figure}
  \centering
  \includegraphics[width=1.0\columnwidth, clip, trim=1cm 14.75cm 10.5cm 1cm]{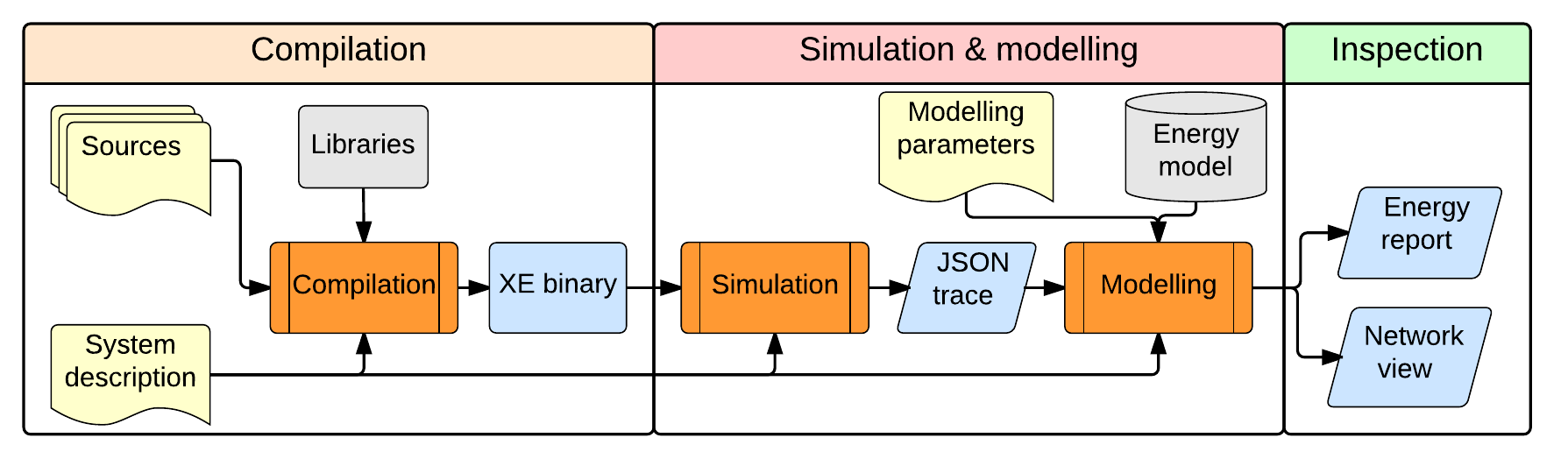}
  \caption{Energy aware multi-core software development workflow.}
  \label{fig:workflow}
\end{figure}

\subsection{Core benchmarks}

The extended core energy model is evaluated using the same benchmark suite as
the original model in~\cite{Kerrison:2015:EMS:2764962.2700104} and on the same
single core \xsonel hardware. This provides direct comparison between the
accuracy of both model versions with respect to the target hardware. The
benchmarks used include the system at idle, various audio sample mixing
variants, operating on multiple independent streams with various levels of
concurrency, one and two concurrent Dhrystone instances, as well as
multi-threaded parallel matrix operations. They are explained in more detail
in~\cite[pp. 20--21]{Kerrison:2015:EMS:2764962.2700104}.

\begin{figure}
  \includegraphics[width=1.0\columnwidth, clip, trim=0cm 2cm 0cm 2cm]{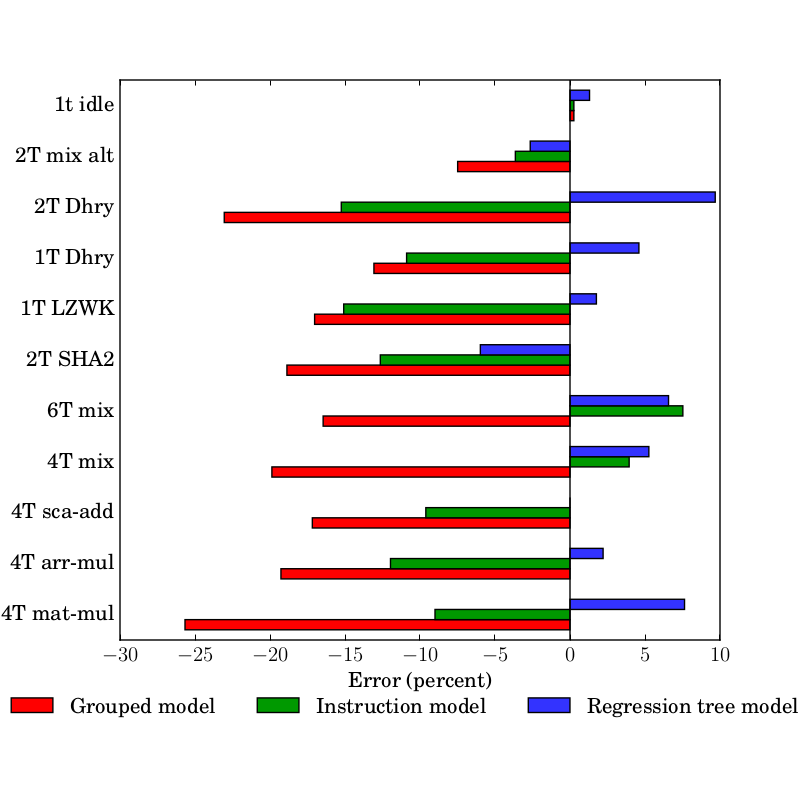}
  \caption{Error of previous (grouped, instruction) models versus new
  (regression tree) model.}

  \label{fig:coreerror}
\end{figure}

\Cref{fig:coreerror} shows the model error for each benchmark with respect to
the hardware measured energy consumption. The regression tree model performs
better than both of the previous model versions in the majority of benchmarks.
Where the original instruction model out-performs the regression tree model,
the difference is approximately a single percentage point of error. The average
and standard deviation of the errors is summarised in
\Cref{tab:coremodelsummary}, where it is evident that the regression tree model
improves overall accuracy whilst reducing variance and the overall range of
error across benchmarks.

\begin{table}
    \centering
    \begin{tabular}{|*{3}{r|}}
        \hline
        \textbf{Model version} & \textbf{Error (\%)} & \textbf{Std. dev. (\%)}
        \\ \hline
        Grouped & $-16.42$ & $6.91$
        \\ \hline
        Instruction & $-7.23$ & $7.45$
        \\ \hline
        Regression tree & $2.67$ & $4.40$
        \\ \hline
    \end{tabular}
    \caption{Geometric mean model error and standard deviation of the tested energy models.}
    \label{tab:coremodelsummary}
\end{table}

\subsection{Multi-core benchmarks}

To test the multi-core model, suitable benchmarks must be used. Those used to
test the core model do not lend themselves to multi-core deployment, due to
their structure and limited, if any, use of channel communication. Instead, two new applications are used for multi-core benchmarking. These are a Finite Impulse Reponse filter (FIR), and the Infinite Impulse Response (IIR) Biquad filter, which will be referred to \texttt{fir} and \texttt{biq} respectively.

Both of these benchmarks are used for applying signal processing, in this case
to streams of audio samples. This is a typical application area for the target
processor, and so a good benchmark selection. Both benchmarks feature multiple
stages, \texttt{fir} implementing seven taps and \texttt{biq} featuring
seven individual biquads.

The concurrent implementation of these applications represent each stage as a
thread, with the progressively filtered samples passed over channels between
threads. The seven threads can be allocated onto a single core the \slicekit,
or distributed across the device's two cores. We test both $7:0$ and $4:3$
thread distributions between the two cores, giving consideration to the fact
that both thread and processor instruction throughput are optimal when a core
has four active threads. We also implement a poorly allocated version of
\texttt{biq} where threads communicate between cores sub-optimally.

\begin{figure}
  \centering
  \includegraphics[width=1.0\columnwidth]{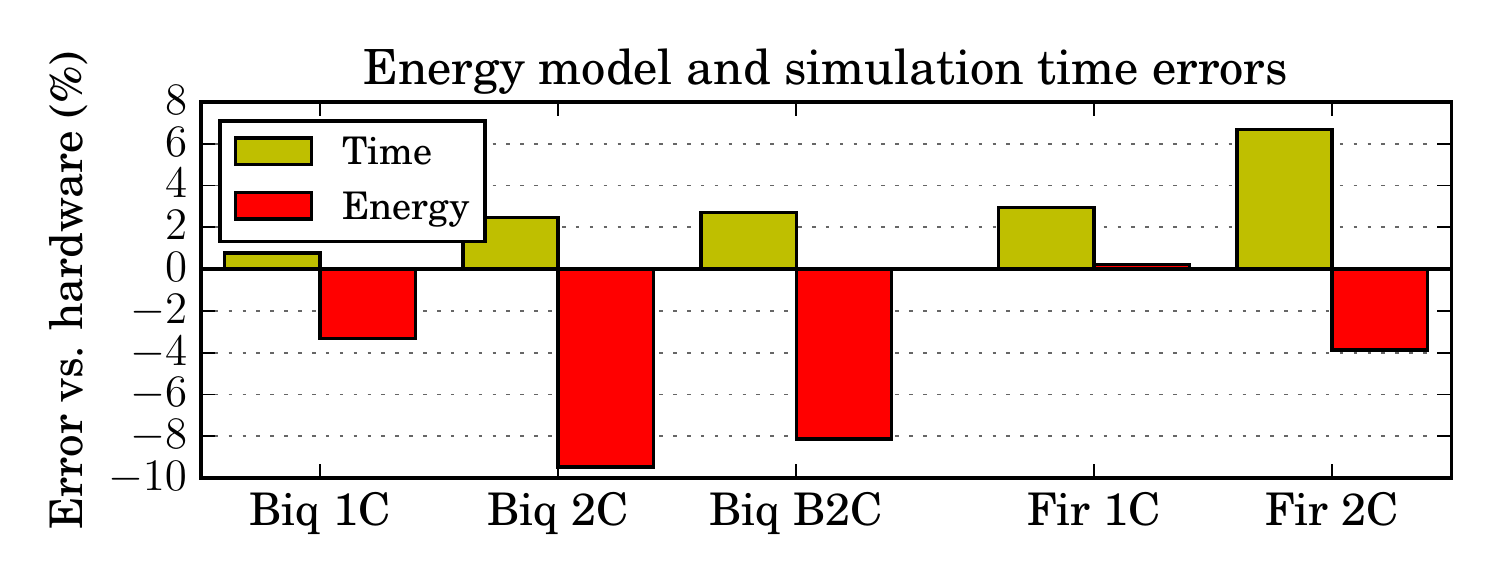}
  \caption{Time and energy errors for \texttt{fir} and \texttt{biq} benchmarks in single (1C) dual core (2C) and for \texttt{biq}, bad dual core (B2C) configurations.}
  \label{fig:mcmodelerror}
\end{figure}

In \Cref{fig:mcmodelerror} the energy and timing errors for the benchmarks are
presented. We observe that in all cases, the simulation over-estimates
execution time, but by less than \SI{7}{\percent}. the energy model under
predicts in the majority of cases, but remains within a \SI{10}{\percent} error
margin.

The overall results are summarised in \Cref{tab:mcmodelavgerror}, which
demonstrates single-digit percentage mean and standard deviations for the
errors. Note that the timing over-prediction counteracts the energy model
under-prediction. Improving one in isolation may in fact reduce the visible
accuracy of the process overall. It is therefore essential to examine the
multiple dimensions of error that are present, in order to direct effort
appropriately.

\begin{table}
  \centering
  \begin{tabular}{|*{3}{r|}}
    \hline
    \textbf{Property} & \textbf{G. mean (\%)} & \textbf{Std. dev. (\%)}
    \\ \hline
    \textbf{Time} & 3.10 & 2.16
    \\ \hline
    \textbf{Energy} & -4.92 & 3.92
    \\ \hline
  \end{tabular}
  \caption{Geometric mean and standard deviation of timing simulation and energy model errors for all \texttt{biq} and \texttt{fir} benchmarks.}
  \label{tab:mcmodelavgerror}
\end{table}

\begin{figure}
  \centering

  \subfigure[Single-core]{
    
    \includegraphics[width=0.9\columnwidth,clip,trim=2.3cm 11cm 2.0cm
10cm]{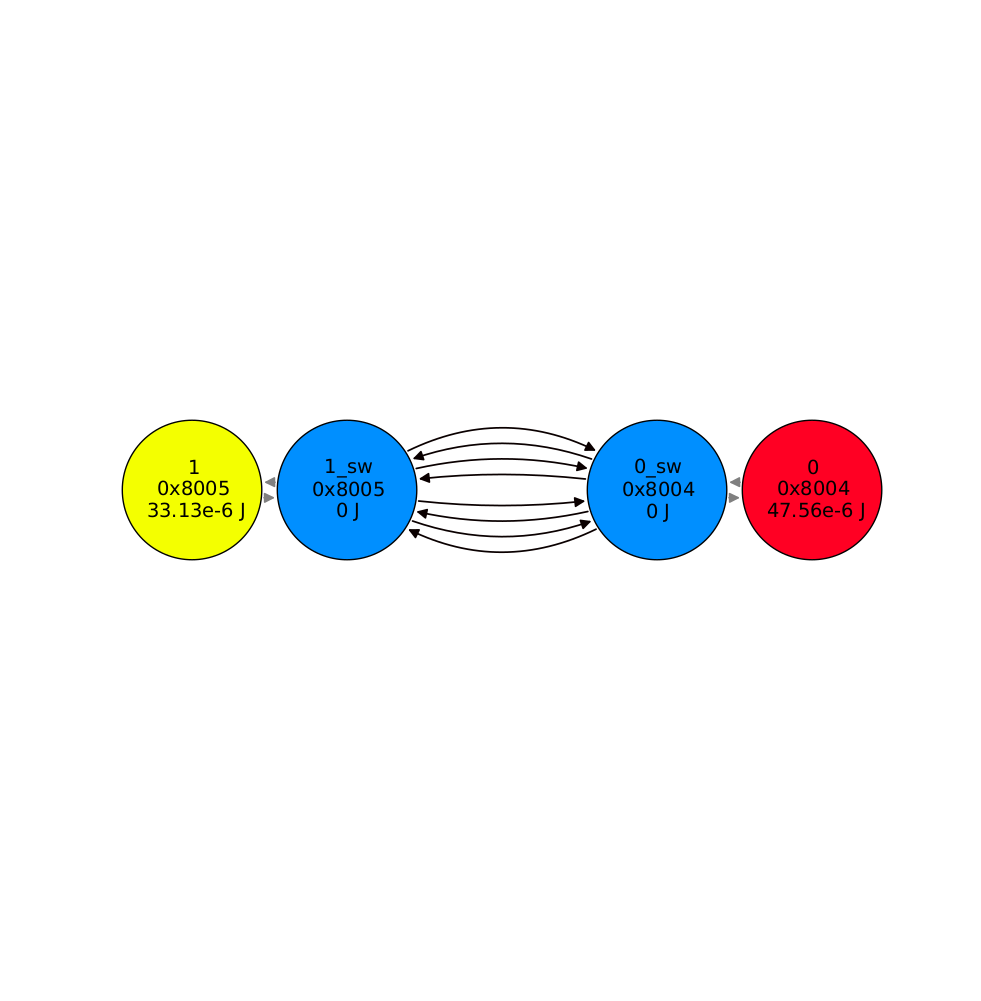}
    \label{fig:1-core}

}

  \subfigure[Dual-core]{
    
    \includegraphics[width=0.9\columnwidth,clip,trim=2.3cm 11cm 2.0cm
    10cm]{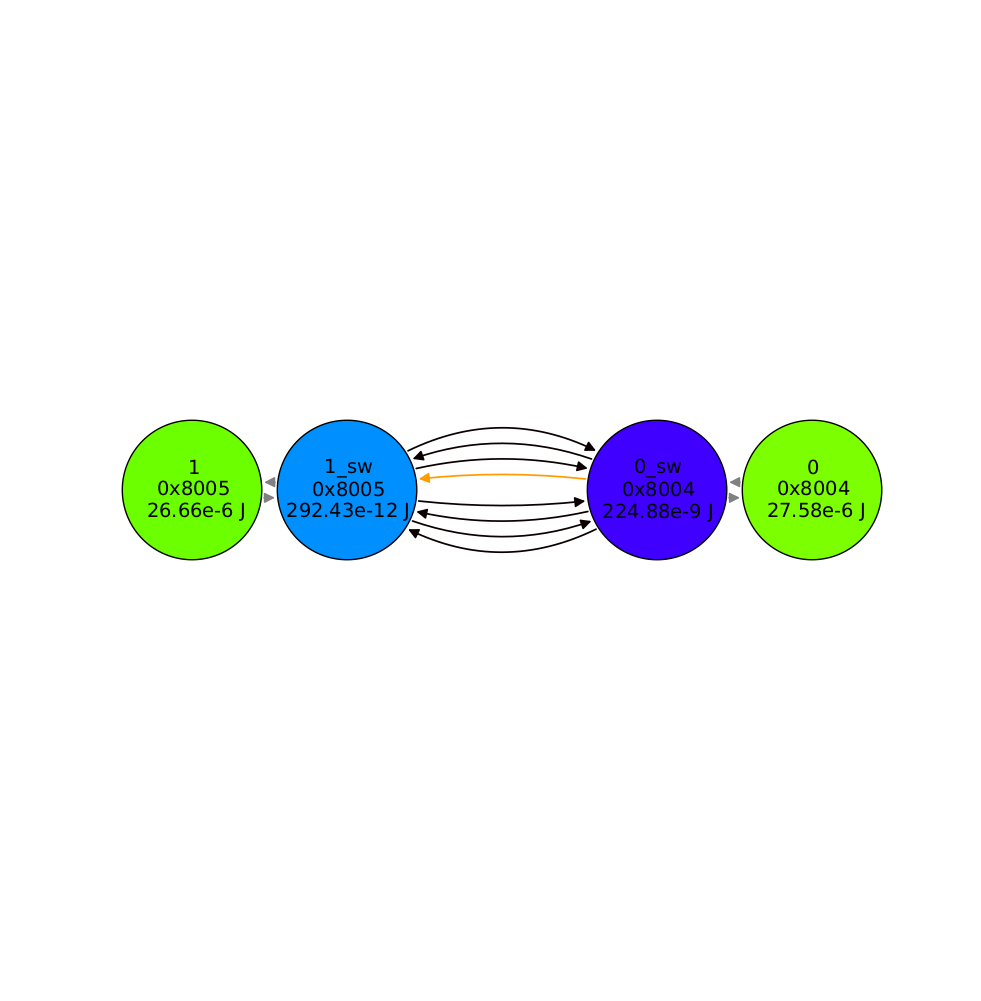}
    \label{fig:2-core}

}

  \subfigure[Dual-core with poorly allocated threads]{
    
    \includegraphics[width=0.9\columnwidth,clip,trim=2.3cm 11cm 2.0cm
    10cm]{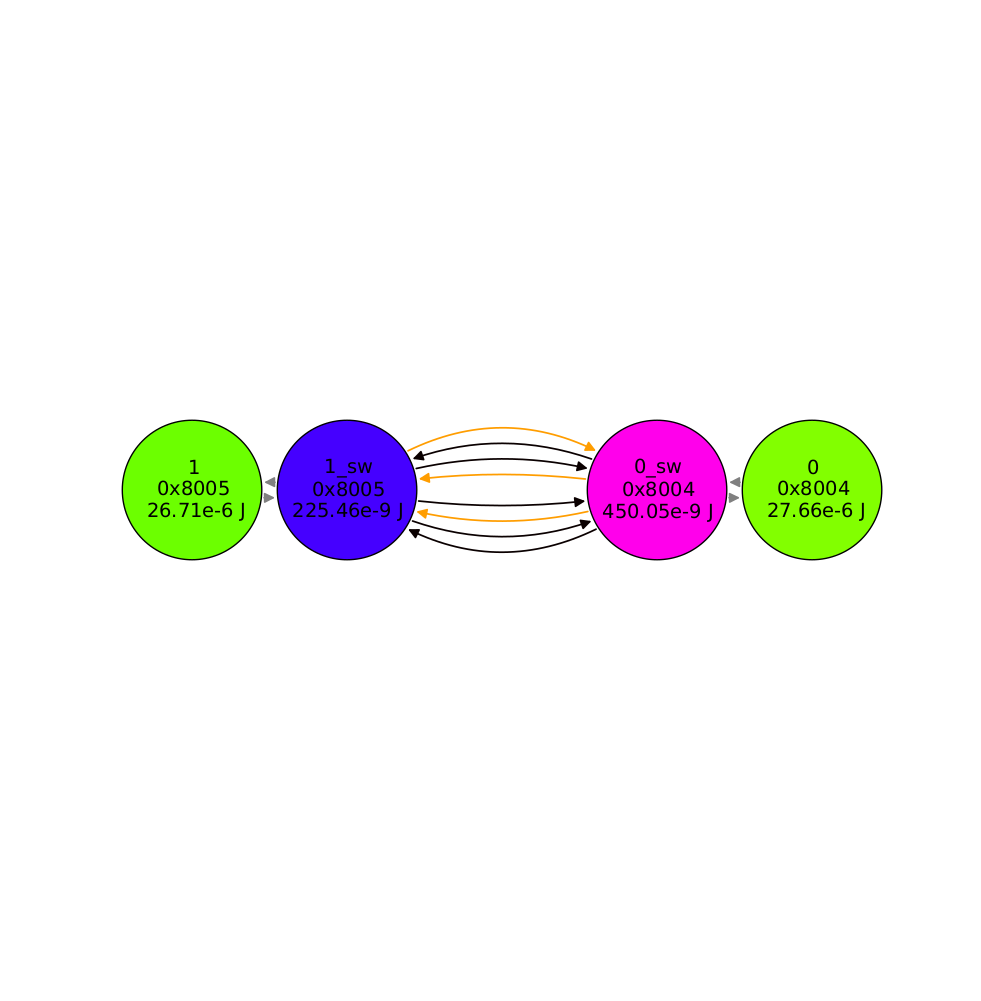}
    \label{fig:2-core-bad}

}
  \caption{Network level energy consumption visualisations the  \texttt{biq} benchmark.}
  \label{fig:mcmodelvisualisation}
\end{figure}

\Cref{fig:mcmodelvisualisation} shows a visualisation of energy consumption in
the cores, switces and interconnect of the \slicekit. Cores are annotated with
the modelled energy consumption, and switches show their energy consumption as
well as the aggregated energy consumed on outbound links. Links are also
coloured by energy. The colouring of the graphs has been scaled to be directly
comparable. Hot/cold are represented as pink/blue for switches and green/red
for cores. Links turn orange as they consume more energy. Only links between
switches are energy modelled, as the core to switch links are captured
implicitly within the core model.

Although visualisation is a less precise representation, it does allow for
comparison and inspection in order to determine \emph{where} energy is
consumed. From these examples, we see that the single core implementation
in \cref{fig:1-core} is the least efficient, taking more energy on the active
core, and resulting in significant energy consumption from leakage in the
otherwise idle core. In \cref{fig:2-core}, the benchmark completes quicker and
work is distributed, so the cores consume less total energy. The communication
cost is insignificant in comparison; some three orders of magnitude less.
Finally, \cref{fig:2-core-bad} is again dual core, but allocates the software
pipeline stages poorly, resulting in three times more core-to-core
communication. The cores consume slightly more energy due to a marginally
longer run-time, and the network cost is three times higher. Not only does this
demonstrate the desirability of distributing the workload across the available
cores, it also demonstrates that energy inefficiency can be introduced with
minimal timing impact, where communication latency may be hidden.

\section{Conclusions and future work}
\label{sec:conclusions}

In this work, a single core, multi-threaded energy model is presented with an average error of less than \SI{5}{\percent}. This is enabled through both instruction set simulator enhancements and a regression tree approach to modelling instructions that cannot be directly energy profiled.

A multi-core model is then described and tested, again supported by
instruction set simulation enhancements. The timing error of the simulator is
shown to be within \SI{7}{\percent} and the energy estimation error within
\SI{-10}{\percent} for two multi-core audio filtering benchmarks, with average
errors of \SI{3.10}{\percent} and \SI{-4.92}{\percent} respectively.

This combination of tools and the demonstrated workflow allows for multi-threaded, multi-core software design space exploration, in order to establish which software definable properties, such as thread allocation and communication patterns, impact the energy consumption of a target device.

A voltage and frequency scaling adaptation of the core energy model is also presented, with a mean squared error of \SI{2.6}{\percent}. In future work, we propose that the \texttt{axe} simulation tool can be further improved to support frequency selection, allowing instruction set simulation, VFS-aware energy modelling, and thus design exploration including VFS parameters.

Additional opportunities for future work include incorporating these new models
into static analysis techniques. Some static analysis techniques have been
demonstrated against prior work on multi-threaded energy
models~\cite{isa-energy-lopstr13,grech15,kyriakosWCEC}, and so there is a
strong case for extending such work to the multi-core level. Simulation based
modelling could also be further extended by examining larger systems, such as
the many-core Swallow system~\cite{swallowOverview}, which assembles
potentially hundreds of XS1-L processors into a compute grid. However,
appropriately sized benchmark applications, or compositions of smaller related
tasks, would need to be identified , adapted or developed in order to exercise
the model and such a system in an appropriate context.

\renewcommand*{\bibfont}{\small}
\printbibliography

\end{document}